\documentclass[conference]{IEEEtran}
\IEEEoverridecommandlockouts
\usepackage{cite}
\usepackage{amsmath,amssymb,amsfonts}
\usepackage{algorithmic}
\usepackage{graphicx}
\usepackage{textcomp}
\usepackage{xcolor}
\usepackage{comment}
\usepackage{soul}

\def\BibTeX{{\rm B\kern-.05em{\sc i\kern-.025em b}\kern-.08em
    T\kern-.1667em\lower.7ex\hbox{E}\kern-.125emX}}
\begin{document}

\title{Practical Issues and Challenges in CSI-based Integrated Sensing and Communication
%\thanks{\IEEEauthorrefmark{4} Daqing Zhang is also with Telecom SudParis, Institut Polytechnique de Paris, Evry Cedex, France.}
%\thanks{\IEEEauthorrefmark{5} Kai Niu is also with Beijing Xiaomi Mobile Software Co., Ltd., Beijing, China.}
\thanks{This work was supported by NSFC under Grant 62061146001, 62172394, PKU-NTU collaboration project, and the Project funded by China Postdoctoral Science Foundation under Grant 2021TQ0048.}
}

\author{
\IEEEauthorblockN{Daqing Zhang\IEEEauthorrefmark{1}\IEEEauthorrefmark{4}, Dan Wu\IEEEauthorrefmark{1}, Kai Niu\IEEEauthorrefmark{1}, Xuanzhi Wang\IEEEauthorrefmark{1}, Fusang Zhang\IEEEauthorrefmark{1}\IEEEauthorrefmark{2}, Jian Yao\IEEEauthorrefmark{3}, Dajie Jiang\IEEEauthorrefmark{3}, Fei Qin\IEEEauthorrefmark{3}}
\IEEEauthorblockA{\IEEEauthorrefmark{1}School of Computer Science, Peking University, Beijing, China}
\IEEEauthorblockA{\IEEEauthorrefmark{2}State Key Laboratory of Computer Sciences, Institute of Software, Chinese Academy of Sciences, Beijing, China}
\IEEEauthorblockA{\IEEEauthorrefmark{3}vivo Mobile Communication Co., Ltd., Beijing, China}
\IEEEauthorblockA{\IEEEauthorrefmark{4}Telecom SudParis, Institut Polytechnique de Paris, Evry Cedex, France}
\IEEEauthorblockA{
Email: dqzhang@sei.pku.edu.cn, \{dan, xjtunk\}@pku.edu.cn, xuanzhiwang@stu.pku.edu.cn,\\ fusang@iscas.ac.cn, \{yaojian.txyj, jiangdajie, qinfei\}@vivo.com}
}

\maketitle

\begin{abstract}
Next-generation mobile communication network (i.e., 6G) has been envisioned to go beyond classical communication functionality and provide integrated sensing and communication~(ISAC) capability to enable more emerging applications, such as smart cities, connected vehicles, AIoT and health care/elder care. Among all the ISAC proposals, the most practical and promising approach is to empower existing wireless network~(e.g., WiFi, 4G/5G) with the augmented ability to sense the surrounding human and environment, and evolve wireless communication networks into intelligent communication and sensing network~(e.g., 6G). In this paper, based on our experience on CSI-based wireless sensing with WiFi/4G/5G signals, we intend to identify ten major practical and theoretical problems that hinder real deployment of ISAC applications, and provide possible solutions to those critical challenges. Hopefully, this work will inspire further research to evolve existing WiFi/4G/5G networks into next-generation intelligent wireless network (i.e., 6G).
\end{abstract}


\section{Introduction}
Next-generation mobile communication network~(i.e., 6G) has been envisioned to go beyond classical communication functionality and provide integrated sensing and communication~(ISAC) capability to enable more emerging applications, such as smart cities, connected vehicles, AIoT and health care/elder care~\cite{liu2021integrated}. Among all the ISAC proposals for 6G network, one practical and promising approach is to leverage the existing wireless devices~(e.g., WiFi, 4G/5G) and explore the sensing mechanisms and applications for the surrounding human, objects and environment within the deployment area of the network. 

Fortunately, significant progress has been made in last decade to employ ubiquitous wireless communication signals such as WiFi/4G/5G for human and environment sensing~\cite{zhang2017toward}. Major efforts have been devoted to develop wireless sensing theories and techniques using Channel State Information~(CSI) for both WiFi and 4G/5G networks due to their ubiquitous deployment, a lot of prototype applications ranging from vital sign monitoring, human computer interaction, human activity recognition, to indoor localization and tracking have been built in both industry and academia~\cite{wanghaoRespiration2016,wu2020fingerdraw,rtfall2017,indotrack2017}. Despite the unique advantages brought over by wireless sensing using WiFi/4G/5G networks such as low cost, ubiquitous infrastructure and non-intrusiveness, a number of problems and challenges are yet to be solved to make the WiFi/4G/5G sensing applications commercially viable and ubiquitous, as they were designed for wireless communications but not for sensing applications~\cite{SurveyZhouGang,zeng2019farsense}. In fact, the current WiFi/4G/5G sensing and communication functionality have very different requirements. How can we achieve integrated sensing and communication~(ISAC) using existing WiFi/4G/5G wireless networks? What are the problems and challenges which hinder the widespread deployment of ISAC systems?   

This paper intends to provide a comprehensive picture of this emerging field by presenting the CSI measurement mechanisms of WiFi/4G/5G wireless networks, identifying the major practical and theoretical problems that hinder the real deployment of ISAC applications, and providing the effective or possible solutions to those critical challenges in the context of WiFi/4G/5G wireless networks. Hopefully, this work will inspire further research to evolve existing WiFi/4G/5G networks into intelligent communication and sensing networks~(e.g., B5G and 6G).

\section{CSI-based Sensing Basics}
% 基于WiFi/4G、5G信号感知都采用CSI。CSI是什么 ？怎么产生的 ？因为CSI这样产生，带来那些感知问题 ？理想CSI怎么拿到 ？
Channel state information characterizes how the wireless signal propagates from the transmitter~(Tx) to the receiver~(Rx) in WiFi/4G/5G networks. It's used by equalizers at receiver so that the fading effect and/or co-channel interference can be removed and the original transmitted signal can be restored for communication purpose.

\subsection{WiFi CSI}\label{sec:wifi_csi}
WiFi Network Interface Controllers (NICs) record variations in the wireless channel using CSI, which captures the combined effects of reflecting, scattering and fading with distance. 
Given a WiFi channel with a central frequency $f$, the CSI~($H(f,t)$) of this channel at time $t$ can be expressed as below:
%$H(f,t) = Y(f,t) / X(f,t)$
\begin{equation}
    \label{eq:howToCalculateCSI}
    \begin{split}
        H(f,t) = Y(f,t) / X(f,t), 
    \end{split}
\end{equation}
where $X(f,t)$ and $Y(f,t)$ are the frequency domain representations of the transmitted and received signals, respectively.

%Although CSI is not reported by all off-the-shelf WiFi cards, fortunately, Atheros-CSI-Tool~\cite{AtherosCSITool} and CSITool~\cite{CSITool2011} enable extraction of detailed PHY wireless communication information~(including the CSI, etc.) from the Atheros WiFi card and Intel 5300 card, respectively.
To measure CSI, The WiFi transmitter sends Long Training Symbols~(LTFs), which contain predefined symbols $X(f,t)$ in the packet preamble. When the LTFs arrive at the receiver, the receiver estimates the CSI based on Equation~\ref{eq:howToCalculateCSI} by using the predefined signal $X(f,t)$ and received signal $Y(f,t)$. 
Note that CSI measurement is done across OFDM subcarriers for a received 802.11 frame.

Unfortunately, for WiFi devices, as the transmitter and receiver are separated and not time-synchronized, there is a time-varying random phase offset $e^{-j\theta_\mathrm{offset}}$ in each measured CSI sample as follows:
\begin{equation}\label{eq:CSIInreality}
    %\begin{split}
    \begin{aligned}
        \widehat{H}(f,t) &= e^{-j\cdot\theta_\mathrm{offset}}H(f,t)\\
        &=e^{-j\cdot(2\pi \Delta f t+\theta_{l}+\theta_{s}+\theta_{d})}H(f,t),
    \end{aligned}
    %\end{split}
\end{equation}
where $\Delta f$ is Central Frequency Offset~(CFO), $\theta_{l}$, $\theta_{s}$ and $\theta_{d}$ are phase errors introduced by Phase Locked Loop~(PLL), Sampling Frequency Offset~(SFO) and Packet Boundary Detection~(PBD) uncertainty, respectively. The amplitude of CSI fluctuates due to automatic gain control~(AGC) for communication as well.

According to the types of source frames, WiFi CSI can be estimated under the following three modes:
\begin{itemize}
    \item \textbf{Beacon frame}. 
    Beacon frame contains all the information about the network which serves to announce the presence of a wireless LAN. 
    Since beacon frames are transmitted periodically, the CSI collected from beacon frames by the receiver has an even sampling interval. The problem is that, the sampling rate of CSI measurement in the beacon frame is too low for most sensing applications.
    
    \item \textbf{Injection frame}.
    WiFi devices can work in monitor mode, so that any communications in a selected channel could be received by those devices. In this mode, a transmitter injects handcraft frames repeatedly with a hardcoded MAC address, and receivers estimate CSI for incoming frames with the specific MAC address.
    Compared to obtaining CSI from beacons, injection monopolizes a channel for sensing, so the CSI measurements are more controllable. For example, in injection mode, researchers can adjust the transmitting interval, the number of frames to be sent, and the transmitting antennas, etc.
    However, injection affects the CSMA/CA of WiFi protocols, which will interfere with the communication of existing devices.
    
    \item \textbf{Data frame}.
    CSI can also be measured from communication data frames while a WiFi device works in AP mode, client mode and IBSS mode. The benefit is that data communication and CSI collection coexist.
    However, as communication happens only when needed, the CSI is sampled with the advent of data frames. Specifically, the CSI sampling interval is uneven due to the random nature of data communication, and CSI samples are insufficient while data communication is idle.
    
\end{itemize}

\subsection{5G NR CSI}
5G NR~(New Radio) CSI is a mechanism that a user equipment~(UE) measures various radio channel quality and reports the result to 5G Base Station~(gNB). 
%In addition to measuring CSI using the predefined symbols and the received signal, the 4G/5G system can also use the random information in data packages to measure CSI, which is different from WiFi system.
%Specifically, the receiver first demodulated the information contained in the data packet, and then took the information as the predefined signal $X(f,t)$, combined with the received signal $Y(f,t)$, to calculate CSI according to Equation~\ref{eq:howToCalculateCSI}. However, when the channel interference is large, the CSI cannot be measured correctly for communication and sensing on account of the wrong demodulated signal. What's more, uneven sampling interval exits due to the nature of random communications.
Compared to WiFi counterpart, there is still no CSI extraction tool that can provide the CSI matrix\footnote{It is worth noting that CSI matrix refers to the matrix of channel estimation results.} to users. Usually, researchers can only use customized hardware devices or software-defined radio platforms to measure CSI on both sides of UE~(downlink) and gNB~(uplink).

%Taking 5G NR as an example, theoretically, 
CSI matrix can be obtained by measurement of reference signal, Synchronization Signals/PBCH block~(SSB) or data packet, etc. The details are presented below:
%According to the types of source packets, 4G LTE/5G NR CSI can be collected under the following three modes:
\begin{itemize}
    \item \textbf{Reference signal}. 
    Reference signals consist of uplink reference signals and downlink reference signals.
    There are several types of downlink reference signals including Channel State Information Reference Signal~(CSI-RS), De-Modulation Reference Signal~(DMRS), Phase-Tracking Reference Signal (PTRS), and Tracking Reference Signal (TRS),  etc. 
    %CSI-RS can be used for CSI measurement, beam management, and time and frequency tracking, etc. Also, the sampling rate of CSI can be controlled, since gNB can allocate time and frequency resources flexibly.
    As a downlink only signal, the CSI-RS a UE receives is used to estimate the channel and report channel quality information back to the gNB. Also, the sampling interval of the CSI matrix is controllable, since gNB can allocate CSI-RS time and frequency resources flexibly.
    DMRS is used to measure CSI for demodulation of the associated data. Therefore, DMRS is always sent with downlink data packets. Due to the randomness of data packets arrival and the uncertainty of time and frequency resources scheduled for data packets, the time and frequency resources for DMRS may be uneven. 
    
    For the uplink, gNB can obtain the uplink CSI by detecting the reference signals such as Sounding Reference Signal~(SRS), DMRS, or PTRS sent by UE.
    %Therefore, DMRS may not meet the resolution requirements for some sensing services.
    
    \item \textbf{Synchronization Signal/PBCH block~(SSB)}. 
    SSB is an ``always-on'' signal which serves devices to find a cell while entering a system. Researchers can obtain CSI measurements from SSB at a fixed sampling interval~(5ms, 10ms, 20ms, 40ms, 80ms or 160ms). 
    %During one periodicity, SSBs are transmitted in different beams at different symbols. 
    Synchronization signal in frequency domain consists of 127 consecutive subcarriers. Assuming the  subcarrier spacing is 15khz, the bandwidth of synchronization signal is less than 2MHz.
    Due to its limited bandwidth, the sensing resolution may not meet the requirements of some sensing services. 
    
    \item \textbf{Data packet}. 
    The CSI can also be measured by data packets.
    Specifically, since the data are unknown in advance for receiver, the receiver first demodulates data from the received signal and then employs the demodulated data to estimate the CSI. Hence, this method is affected by the data demodulation performance, i.e., the data need to be demodulated correctly, otherwise demodulation errors will affect the estimation of the channel matrix information. 
    Besides, compared to the reference signal that is dedicated to measure CSI, the uncertainty of data sent by users results in the biased auto-correlation and inter-correlation calculation, which may further degrade the channel estimation performance. 
    %For the monostatic sensing mode with echo detecting, the data signal is known at the receiver, and the channel matrix can be estimated directly using the known data signal. 
    %However, since the data signal is different from the dedicated sequence used for the reference signal, the autocorrelation and intercorrelation properties of the data signal cannot be guaranteed, and the channel estimation performance may be degraded. 
    %What's more, uneven sampling interval exits due to the nature of random communications.
    %However, when the channel interference is large, the CSI cannot be measured correctly for communication and sensing on account of the wrong demodulated signal. What's more, uneven sampling interval exits due to the nature of random communications.
    
\end{itemize}

%Taking 5G NR downlink as an example, theoretically, the channel matrix (equivalent to CSI for wifi) of NR system can be obtained by UE measurement of downlink reference signals, SSB (Synchronization Signal and PBCH block) or data signals, etc. 

%CSI parameters are the quantities related to the state of a channel. The user equipment~(UE) reports CSI parameters to the access network node~(gNB) as a feedback. 
%The CSI includes several parameters, such as the Channel Quality Indicator~(CQI), the Precoding Matrix Index~(PMI), and the rank indicator~(RI). 
%Upon receiving the CSI parameters, the gNB schedules downlink data transmissions (such as modulation scheme, code rate, etc.) accordingly. 

\section{Problems and Challenges in CSI-based ISAC}

\subsection{Practical Problems}\label{sec:practical_problem}

CSI is designed for communication purposes, it is not optimized for sensing. Based on our experience in developing CSI-based sensing systems, several practical challenges need to be addressed.

\subsubsection{Accessibility of CSI}
Each OFDM-based wireless communication system estimates CSI for channel equalization. However, until today, few commercial WiFi/4G/5G devices make CSI available except the early release of the experimental CSItools for Intel 5300 WiFi chips \cite{CSITool2011} and Atheros Ath9k series WiFi chips \cite{AtherosCSITool}. While these CSI extraction tools are available for research purposes, the commercial WiFi CSI tools from the major WiFi chip manufacturers such as Intel, Broadcom and Qualcomm are still proprietary and not open for public access.

Compared to the WiFi counterpart, there is still no CSI extraction tool that can work on commodity 4G/5G chips. SrSLTE \cite{gomez2016srslte}, an open-source 4G LTE implementation on the USRP software radio platform, can be modified to extract CSI value for research purposes. It is one prototype platform that researchers can study CSI-based sensing on 4G LTE/5G communication systems~\cite{Chen2020RobustDH}.

Since the channel estimation and equalization algorithms for most communication systems are implemented on-chip, without the help of chip designers, it is still a problem to access CSI and design commercial ISAC products for WiFi/4G/5G systems.

\subsubsection{CSI sampling}
Due to the requirement of persistent sensing and the nature of random communications, the CSI estimation rate in wireless communication differs from that required by the CSI-based sensing system, leading to the second important issue in ISAC design.
Depending on which mode that CSI estimation is based on in Sec. \ref{sec:wifi_csi}, WiFi CSI sampling may suffer from problems of low sampling rate~(extracting CSI from AP beacon) or uneven sampling interval~(extracting CSI from data frames).

In the 4G LTE/5G NR network, all UEs in a cell are time-synchronized and CSMA/CA is not used. In the FDD network, the downlink or uplink slots/subframes are continuous. Assuming the signals for sensing are transmitted in one symbol every millisecond, the sampling rate is 1000Hz. While in the TDD network, the downlink or uplink slots/subframes are discontinuous due to downlink-uplink transmission periodicity, which has a maximum value of 10ms. Assuming the signals for sensing are transmitted in one symbol every downlink-uplink transmission periodicity~(e.g., 10ms), the CSI sampling rate is 100Hz. 
In a practical network, the CSI sampling rate also depends on the density of reference signals. % For 5G NR, CSI can be obtained by measurement of reference signals, SSB or data signals, etc. 
For DMRS and data signals, the CSI sampling interval is uneven, due to the randomness of data packet arrival and the uncertainty of time and frequency resources scheduled for data packets. For SSB, since it is an ``always-on'' signal, CSI sampling depends on the configured SSB periodicity. For CSI-RS, the sampling rate depends on the periodicity of CSI-RS configured by gNB.

\subsubsection{Unsynchronized transceivers}
The unsynchronized transceiver problem is another severe issue in ISAC systems. As the transmitting and receiving devices in the ISAC system use their hardware clock for timing and carrier generation, the unsynchronized transceiving devices will generate random phase offset in CSI \cite{AtherosCSITool,chen2019residual,ni2021uplink}. The carrier frequency difference between transmitter and receiver renders carrier frequency offset~(CFO). CFO is the same for all the subcarriers, but it is different across packets. The timing difference between transceivers introduces sampling frequency offset~(SFO). SFO has a linear relationship with subcarriers. To make things worse, the phase offset drifts over time. As there is a random phase offset in CSI, the amplitude and phase of CSI are no longer orthogonal to each other. Therefore, it is difficult to reliably sense small motions and accurately extract Doppler frequency shifts.

\subsubsection{CSI distortion}
CSI-based sensing techniques assume to sense human information by monitoring the changes of wireless channels. However, the actual CSI in a communication system includes not only the wireless channel of air transmission but also multiple processing modules at the transmitter and receiver. Therefore, the dynamic adjustment of the parameters such as precoding, beamforming, and automatic gain control in the communication devices will affect CSI measurement. For instance, DMRS in NR can be transmitted with SVD~(Singular Value Decomposition) based precoding, so the result obtained by detecting DMRS may not reflect the original channel information.

The distortion in CSI is manifested at different levels, which depends on the hardware design. For example, the adjustment of AGC may be different between antennas or even across subcarriers, which leads to impulse-like noise in CSI amplitude. 

\subsubsection{Distributed Sensing Data Collection}
In non-contact human sensing, the sensing system is usually composed of a transmitter and multiple receivers located at different locations. Many sensing applications, such as motion tracking and gesture recognition, need to collect CSI from multiple receivers.
Since CSI can only be generated when two transceivers communicate, it is difficult for other non-participating devices to estimate CSI at the same time. Therefore, it is a challenge to design an ISAC system to enable multiple devices to estimate CSI simultaneously.

In addition, even if we can estimate CSI on multiple receivers at the same time, how to effectively transmit all CSI to a central processing server in the network is another problem. For example, a WiFi router is a good candidate processing server in a smart home. We hope to aggregate all CSI to the AP router for further processing, regardless of whether the CSI is collected in the AP or WiFi client.

\subsection{Theoretical Problems}

Besides the practical CSI problems, WiFi/4G/5G CSI-based sensing systems also face five theoretical problems.

\subsubsection{Sensing Granularity Limit}
% 感知粒度的极限- 菲涅尔区揭示

The surge of work in CSI-based wireless sensing over the last few years demonstrates the powerful sensing ability of ubiquitous wireless signals such as WiFi and 4G/5G. Especially, WiFi CSI signals can be used to monitor the millimeter-level chest movement of human respiration ~\cite{wisleep2014,liu2015tracking,wanghaoRespiration2016,zhangdiffration2018,zeng2019farsense,zeng2020multisense} and detect the centimeter-level displacement of moving fingers~\cite{Wikey2015,niukaiconext2018,wimorse2019,wu2020fingerdraw}. Though promising, one basic issue is what is the minimum displacement that can be detected by the ubiquitous wireless signals. In other words, \textit{what is the sensing granularity limit of these ubiquitous wireless signals}? If there is a theoretical model that can clarify this issue, it will help researchers to recognize the sensing limits of the ubiquitous wireless signals and understand what application can or cannot be supported.

% 大量工作表明表明WiFi具有很强的感知能力，
% 可以感知细粒度的呼吸、手指敲击等厘米级/毫米级的活动
% though promising,  
% 问题是WiFi感知的粒度极限是多大？ 也就是xx。 

\subsubsection{Sensing Boundary Limit}
% 感知边界-包括如何通过SNR度量、最远感知范围、通信距离和感知距离的均衡等
% 知道通信范围在部署通讯设备时至关重要，however， 微小小于通信距离，但是怎么刻画，最远的距离是多大。
When deploying the communication devices into real-life environments, it is very important to know the maximum communication distance between the adjacent devices so that the communication function can work well. For instance, WiFi is designed to communicate with each other within 300 meter range. Different from the communication function that leverages the signals propagating through direct paths, the sensing function employs the signals that are reflected or diffracted by the target. Thus, the sensing coverage is much smaller than the communication coverage due to the long propagation path and low reflectivity of the target. However, in the field of CSI-based sensing, \textit{there lacks a metric to characterize the sensing ability and inform the sensing boundary limit for a given transmitter pair}.

\subsubsection{Position Dependency Issue}
% 感知位置的依赖性- 振幅/相位的位置依赖，速度的位置依赖
%　人的活动是随机的，理想情况下，在所有可能的位置和朝向时实现一致的性能。　然而已有工作发现位置依赖性问题，具体的。　
One key assumption of CSI-based sensing is that there exists a fixed mapping between human activity and CSI signal features, i.e., the corresponding CSI signal features are consistent for the same activity, yet different for different activities. The features can be the amplitude and phase of CSI signals or frequency and speed extracted from received CSI signals. However, the Fresnel Zone model~\cite{wanghaoRespiration2016,wu2016widir,zhang2017toward,zhangdiffration2018,zhangdiffration2019} reveals when the same activity is conducted at different positions~(i.e., locations and/or orientations), the amplitude and phase features of the CSI signals can vary significantly, resulting in unstable activity recognition performance. Furthermore, the frequency and speed features are also proved to be position-dependent~\cite{niugesture2021}. Considering the diversity of application scenarios in real-world environments, \textit{it is important to develop CSI-based sensing mechanisms that are position-independent}.

\subsubsection{Automatic Segmentation for Continuous Activities}
% 感知信号的自然连续切割
% 人的活动是连续的， 例如，走路这跌倒了，如何切割？
CSI-based sensing has enabled lots of real-world applications including gesture recognition~\cite{niugesture2021,gaogesture2021}, activity recognition~\cite{wifall2017,ARAlarm2017}, and localization and tracking~\cite{indotrack2017,wu2021witraj}. However, most of existing prototype systems manually segment the CSI fragments induced by the activities in interest from the received continuous CSI signals or assume there is an obvious pause between the adjacent activities for CSI signal segmentation. In fact, the activities performed by the human target are usually continuous without pause. For instance, the target may walk to a chair and then sit down, in which there is no pause between the walking and sitting activities, so are the received CSI signals. Thus, \textit{how to automatically segment the activities in interest from the continuous activities in daily life is challenging}. 

\subsubsection{Privacy and Security Issue}
% 隐私和安全

As the wireless signals can propagate through the obstacles such as walls, CSI-based sensing can detect/recognize activities in a through-wall manner. However, this also introduces many privacy and security issues. For instance, the malicious hackers and attackers can also leverage the received CSI signals to infer the target's daily activities and issue an attack. Moreover, the information leakage is very difficult to be detected since CSI-based sensing is also non-intrusive. Therefore, \textit{more attention should be paid to the privacy and security issue when leveraging CSI-based sensing to provide services.}

\section{Potential solutions and research directions}

\subsection{Solutions for Practical Issues}

\subsubsection{Accessibility of CSI}
In order to easily obtain CSI from diverse communication hardware of different manufacturers, it is preferable to standardize it as an unified CSI acquisition API. Of course, this requires the efforts from both the chip manufacturers and communication equipment vendors. It is also suggested that the CSI acquisition API can be included in future standard, so that all manufactures can support CSI acquisition. The CSI acquisition API should include key sensing information such as the hardware timestamp, the MAC address of the transmitting device, the MIMO antenna configuration, CSI matrix arranged in an unified order of antenna pairs and subcarriers.

\subsubsection{CSI sampling}
The problems of low CSI sampling rate and uneven sampling interval can be solved in many ways. For WiFi signals, CSI can be extracted by combining beacon frames and data frames. Beacon frames ensure that CSI sampling can be carried out at a lower rate when no communication occurs; while data frames ensure a high CSI sampling rate. Although the interval of data frames is uneven in WiFi, we could interpolate the CSI samples to make it as uniform as possible. It is hoped that future WiFi standard can take into account both the sensing and communication requirements, so that all WiFi receiving devices can estimate CSI at an appropriate sampling rate.

For 4G/5G signals, the sampling uniformity problem is not as serious as WiFi. The first reason is that CSMA/CA is not used in 4G/5G. The second reason is that gNB can flexibly configure the periodicity of some signals, e.g., CSI-RS or SSB to meet the sampling rate requirements.

\subsubsection{Unsynchronized transceivers}
The unsynchronized transceivers produce random phase offset in CSI, which need to be removed before further processing. Existing studies show phase offset can be modeled as linear fitting problem \cite{AtherosCSITool} or spectrum estimation problem \cite{chen2019residual}. By estimating phase error across CSI samples, part of the phase offset could be mitigated. The drawback is that the residual phase offset is still not negligible. CSI-ratio model offers a practical solution to this problem by taking the division operation between the CSI of two receiving antennas to systematically eliminate the phase offset \cite{zeng2019farsense,wu2020fingerdraw,Exploring2020}. CSI-ratio model works well for a single moving target, and a general model is proposed to handle the multiple targets case \cite{zeng2020multisense}. In the future, it is desired to coordinate both the transmitter and receiver, so that we can use a unified solution to eliminate phase offset and achieve device synchronization.

\subsubsection{CSI distortion}
CSI distortion is caused by the dynamic adjustment of the communication modules. Due to different optimization strategies adopted by communication equipment suppliers, CSI distortion performs differently. Therefore, there is no universal solution to mitigating distortion at the software level for different communication devices.

CSI distortion caused by communication modules can be fixed at the hardware level. Since the transmitting and receiving devices have all the information of automatic gain control adjustment, the receiving device can 'undo' the adjustment of CSI value so that the reported CSI does not include various changes introduced by communication hardware. An ideal approach is to standardize it in the CSI acquisition API.

\subsubsection{Distributed Sensing Data Collection}
The collection of distributed CSI data can be achieved by using broadcast or multicast frames in wireless networks. For example, in WiFi, CSI data can be collected simultaneously at multiple receivers by receiving beacon frames or short multicast messages from AP. In 4G LTE/5G NR, CSI can be estimated on multiple UEs using SSB information. Another way to collect distributed CSI data at multiple receivers is to receive unicast frames at each receiver in turn. The difficulty lies in the need to maintain a sufficient CSI sampling rate and a uniform transmission interval at the same time.

Note that there are several ways to collect the distributed CSI data from multiple devices. In the case of unicast, a practical solution is to estimate CSI from a central device~(such as WiFi AP). By sending a query packet to each WiFi client and waiting for the ACK frame, the CSI is then estimated based on the ACK frame \cite{abedi2020wifi}. In the case of broadcast or multicast, one possible idea is to piggyback CSI estimated by the client to the transmitter, such as embedding CSI in each ACK frame. For delay tolerant sensing applications, we can aggregate CSI samples and decrease the piggyback frequency to reduce the channel occupation.

\subsection{Solutions for Theoretical Issues}

\subsubsection{Sensing Granularity Limit}
Zhang et al.~\cite{zhang2017toward, wanghaoRespiration2016, wu2016widir} propose the Fresnel Zone model to depict the mathematical relationship between CSI and target's motion trajectory, target's position with respect to transceivers when the target locates outside the First Fresnel Zone~(FFZ). They further extend the Fresnel Zone model in the FFZ where the diffraction effect dominates~\cite{zhangdiffration2018}. Based on the Fresnel Zone model, why and when the millimeter-level chest movement induced by human respiration can be detected are clarified. Together with the dynamic vector change model~\cite{niukaiconext2018}, the Fresnel Zone model sheds light on the scale of human activities that can be sensed and recognized using ubiquitous wireless signals -- that is, the signal's possible sensing granularity limit.

\subsubsection{Sensing Boundary Limit}

In the communication field, Signal to Noise Ratio~(SNR) is proposed to quantify the communication capability of the received signal. If the SNR is lower than a threshold, communication may fail due to a high bit error rate. Furthermore, the minimum SNR that can be tolerated by the receiver determines the maximum communication range. In CSI-based sensing field, Wang \textit{et al.} propose the Sensing Signal to Noise Ratio~(SSNR) to quantify the sensing capability of the received signal~\cite{sensingCoverage2021}. With the metric of SSNR, both the sensing coverage and the sensing boundary can be determined. 

\subsubsection{Position Dependency Issue}

Though it is challenging to completely solve the position dependency issue, significant progress has been made in developing position-independent applications such as  human respiration monitoring, gesture recognition, and human tracking. For instance, the complementary feature between amplitude and phase of CSI signals is employed to address the ``blind spot'' issue in human respiration monitoring~\cite{FullBreathe2018}. And the relative movement patterns are proposed to achieve position independent gesture recognition~\cite{gaogesture2021}, while a selection scheme is proposed to select the optimal devices to track human target position independently. An unified solution is needed to address the position dependency issue in CSI-based wireless sensing.

\subsubsection{Automatic Segmentation for Continuous Activities}

Automatic segmentation for continuous activities is still an open issue in CSI-based wireless sensing. To our knowledge, very few work has addressed this tough issue. In RT-Fall~\cite{rtfall2017}, a transition-based segmentation method is proposed to automatically segment fall activity in the continuously captured CSI signals.  
%WiBorder~\cite{WiBorder2020} can continuously discriminate the through-wall activity and achieves the real-time intrusion detection in daily environments. 
WiBorder~\cite{WiBorder2020} segments continuous daily activities by identifying still, in-place activity and walking across different rooms.
Only when the continuous activities can be automatically segmented, the CSI-based activity recognition can be realized and widely deployed in real-life scenarios.  

\subsubsection{Privacy and security Issue}

To address the privacy and security issue in CSI-based wireless sensing, one possible way is leveraging a precoding scheme to encrypt the CSI signals. Only the authorized devices can decrypt the CSI signals. Thus even though the attackers can eavesdrop on the CSI outside the walls, they can not decrypt the CSI signals to infer targets' activities. 
% 通过编码加密的方式，使得其他人无法获取到CSI

\section{Conclusion}
In this paper, in order to empower the existing wireless networks (e.g., WiFi, 4G/5G) with augmented sensing ability using CSI, we present the different CSI generation mechanisms in WiFi and 4G/5G wireless networks, identify ten major practical and theoretical problems that hinder real deployment of ISAC applications, and provide possible solutions to those critical challenges. Hopefully, this work will inspire further research to evolve existing WiFi/4G/5G networks into next-generation wireless networks (e.g., B5G and 6G).

\bibliographystyle{IEEEtran}
\bibliography{ref}

\end{document}